\documentclass[journal]{IEEEtran}
\usepackage{amsmath}
\usepackage{graphicx}
\usepackage{subcaption}
\usepackage{amsthm}
\usepackage[utf8]{inputenc}
\usepackage[english]{babel}

\begin{document}
\title{Windowing and random weighting based cubature RTS smoothing for target tracking }

\author{Mundla~Narasimhappa,~\IEEEmembership{ Member,~IEEE,}
\thanks{His with the Department
of Mechanical Engineering, University of Surrey, Surrey,
GU2 7XH, UK e-mail: (see n.mundla@surrey.ac.uk.}%
}

\maketitle

\begin{abstract}

This paper presents windowing and random weighting (WRW) based adaptive cubature  Rauch–Tung–Striebel (CRTS) smoother (WRWACRTS). The Unscented KF (WRWUKF) has already existed as an alternative to nonlinear smoothing solutions. In the proposed method, both windowing and random weighted estimation methods are combined together, and used to estimate the noise statistics. Subsequently, the weights of each window are adjusting randomly, and update the process and measurement noise covariances matrices at each epoch. The developed WRWACRTS algorithm  overcomes the limitation of the conventional CKS. The Lyapunov function-based approach is used to investigate the convergence proof of the WRWACRTS algorithm. A numerical example is shown to demonstrate the performance of the proposed algorithm.

\end{abstract}

\begin{IEEEkeywords}
Adaptive Cubature Kalman, window method, Random weighted method, Rauch–Tung–Striebel (RTS) smoother.
\end{IEEEkeywords}

\IEEEpeerreviewmaketitle

\section{Introduction}

\IEEEPARstart{I}n several practical problems of target  tracking has been an active research area for many wide range application in radar, wireless sensor network,  navigation and sonar are in \cite{bar2004estimation}. Estimating the kinematic states (e.g., position, velocity and acceleration) including in target tracking and is the main objective for a moving object under the the noisy measurements. Target tracking typically involves arrival time, arrival time difference, bearing angle (angle of arrival) and received signal strength, which are all provided by the sensors. Bearings-only target tracking is the main challenging problem that indicates the nonlinear relationship between target dynamics and bearing only angle measurements in the 2D plane. Several state estimation methods have been developed for bearings-only tracking as in the literature, Kalman filter \cite{aidala1979kalman}, Extended Kalman filter (EKF) \cite{toloei2014state},  Unscented Kalman filter (UKF) \cite{wan2000unscented} and cubature Kalman filter (CKF) \cite{arasaratnam2009cubature} have been developed.  The basic idea of the nonlinear estimators  are to approximate mean and covariances of the state. The performance of nonlinear KF depends on the priori knowledge of the system and measurement models. As compared to other estimators, the CKF has better than EKF and the same with UKF for solving the higher order nonlinear systems. Moreover, it can avoid linearization of the nonlinear system. The accuracy of the CKF and UKF are the same, but better than the EKF algorithm \cite{arasaratnam2009cubature}. Because of these merits, the CKF is widely used in control and tracking applications \cite{meiqin2015bearing, zhong2015random}, and used in robotics and Inertial \cite{crassidis2011optimal}, control and guidance navigation  \cite{ali2019application, yu2014slam, caccia2008basic}. \cite{ fang2020noise}.

As mentioned before, the EKF, UKF and CKF are developed based on the idea of forward filtering, whereas in the smoothing estimation, the state is estimated by use of full observation information of each moment. The accuracy of the optimal smoothing algorithm is higher than that of the CKF \cite{arasaratnam2011smoothcubature}. In \cite{sarkka2008unscented}, unscented Rauch-Tung-Striebel smoother (URTSS) and cubature Rauch-Tung-Striebel smoother (CRTSS) \cite{arasaratnam2011smoothcubature} have been developed and  proved that URTSS makes the better estimation accuracy than that of the conventional UKF and CKF, respectively. In practice, priori knowledge of the  system and noise models are unknown and the uncertainties in measurement model may leads to large errors, even the filter becomes diverge  \cite{jia2015new}.

Several nonlinear adaptive CKF (ACKF) have been  developed based on innovation or residual based adaptive estimation (RAE). The innovation or residual vector is a additional information to the filter, used to estimate the noise statistics and, followed by sliding window average method. Random weighting estimation (RWE) is an adaptive method has been developed based on statistics and probability theory \cite{gao2015sage, gao2015windowing}. The RWE method has a simple computation method, unbiased estimation and easy to solve large sample problems. Moreover, noise parameters are updated without considering an exact probability distribution of the state characteristic \cite{Gao:2014} \cite{gao2015sage}. In the RWE analysis, random variables, $ \bf X_1, \bf X_2, ..., \bf X_n$ are independent and identically distributed (i.i.d) observations. $\bf F ({x})$ and $\bf F_{n}(x) $ are the common distribution function and corresponding empirical distribution function of random variables. Let  $\bf x_1, \bf x_2, ..., \bf x_n $ are the sample realization. From which, an empirical distribution function  is calculated \cite{Gao:2014}  as 

\begin{equation}
{\bf F_{n}(x)}={\frac{1}{n}}\sum_{i=1}^{n}\bf I_{(\bf X_{i}< x)}
  \label{stateSpaceForm1}
  \end{equation}
and also $F(x)$ is the random weighting estimation, it can defined as

\begin{equation}
{\bf H_{n}(x)}=\sum_{i=1}^{n}{\bf \lambda_{i}}I_{(\bf X_{i}< x)}
  \label{stateSpaceForm1}
  \end{equation}
where $ \bf I_{(\bf X_{i}< \bf x)}$ is the indicator function, it is defined as
\[
 \bf I_{{X_i}<\bf x}=\left\{ \begin{array}{cc}
    1, & {\bf I_{{X_i}< \bf x}}.\\
    0, & {\bf I_{{X_i}> \bf x}}.
 \end{array} \right.
\]

and $\ [\bf \lambda_{1}, \bf \lambda_{2},.......... \bf \lambda_{n} \ ]$ is the random vector. It follows the Dirichlet distribution D(1, 1, ..., 1),
and that is $ \sum_{i=1}^{n}{\bf \lambda_{i}}=1$.   The joint density function of $\ [\bf \lambda_{1},\bf \lambda_{2},.......... \bf \lambda_{n} \ ]$ is $ \bf f(\lambda_{1},\lambda_{2},..........\lambda_{n})={\bf \tau_{n} }$. where $(\bf  \lambda_{1}, \bf\lambda_{2},.......... \bf \lambda_{n} )  \in  \bf  D_{n}$ and $ \bf D_{n-1}=\{ [\bf \lambda_{1}, \bf \lambda_{2},.......... \bf \lambda_{n} : \bf \lambda_{k} > 0 \ (k=1,2,...n-1),\  \sum_{i=1}^{n-1}{\bf \lambda_{k}}< 1\} $.

Recently, the windowing and random weighted estimation (WRWE) theory was explored in the nonlinear UKF \cite{gao2015sage}. To the best of authors knowledge, there has been very limited research regarding the use of WRWE in ACKF for adjusting the filter parameters and in case of application too. The RWE based cubature KF as been  developed  with bias estimator in \cite{gao2019adaptively} for improving the accuracy of nonlinear dynamical system. We extend the same work into adaptive cubature RTS smoothing with noise statistics estimators.

This paper presents windowing and random weighting-estimation (WRWE) based adaptive Cubature RTS smoothing (WRWACRTS). In the proposed method,  WRWE method is combined together, and used to adjust the weights randomly at each window for updating the noise statistics. Moreover, presenting the convergence proof of the WRWACRTS algorithm. To demonstrate the performance improvement of the proposed WRWACRTS algorithm with numerical example.

The rest of the paper was organized as follows; The related work is studied in Section II. The description of the traditional cubature smoothing filter is presented in Section III. Section IV provides the proposed algorithm.   Numerical examples along with performance analysis of the developed algorithm is given in Section IV. Numerical simulation results are demonstrated in Section V.   Section VI conclusions of the paper.

\section{Related Work}
When the system and  noise models are not known exactly in practice, even-though the system models are nonlinear behaviour.  Over years, extended KF (EKF),  Unscented Kalman filter (UKF) and cubature Kalman filter (CKF) have been developed for estimating the state of a nonlinear dynamical system\cite{leong2013gaussian}.
In the EKF, Jacobin/Taylor series can be used to approximate the  mean and covariance of state vector. The performance of the EKF is limited by the Jacobin matrices calculation that leads large errors. To do this,  UKF has been developed based on Unscented Transform (UT) \cite{wan2000unscented}, for addressing the EKF limitations. In practice, system and noise models are vary with time. Moreover, the UKF accuracy is limited for higher order system analysis. Alternatively,  CKF has been developed for solving the higher order nonlinear system analysis. However, the computation load  and accuracy of CKF and UKF are similar, but the filter converges slowly.

Aforementioned literature, optimal smoothing algorithms can be divided into (i) two-filter smoother and (ii) the Rauch–Tung–Striebel (RTS) smoother \cite{arasaratnam2011smoothcubature}. In the two-filter smoother, combined two filters and run them in both forward and backward direction. The performance of two filters smoother can provide poor results in practice \cite{arasaratnam2011smoothcubature}. In RTS smoother \cite{sarkka2008unscented, sarkka2010gaussian}, forward filter and backward smoother were combined both algorithms and widely uses in practical analysis. Recently,  RTS type of CKS has been developed based on high-degree  of the spherical and the radial rules. In this analysis, Genz's and  Mysovskikh's methods have been used to generate the number of spherical points and update the weights too \cite{jia2015new}.
In recently, several nonlinear smoothing algorithms have been developed on  extended RTS (ERTS) and  unscented Rauch-Tung-Striebel smoother (URTSS) \cite{wang2020maximum, sarkka2008unscented} for smoothing characteristics of a nonlinear systems. The estimation accuracy of URTSS is much better than that of the EKF and UKF. In addition, there are other smoothing filter developed with central difference RTS (CDRTS) smoothing \cite{psiaki2005backward}, cubature RTS (CRTS) smoothing algorithm \cite{xia2015state,ali2019application} also have been developed for nonlinear systems. As Compared with the URTS and CDRTS, CRTS achieves the same accuracy level, but higher numerical stability. In the later development, moment matching method has been utilized to generate the radial rule points and weights with arbitrary degree of accuracy. Although the third-degree CKS works well but  when dynamical system has high non-linearity and large uncertainty \cite{gao2019adaptively, meiqin2015bearing, wang2020maximum}, it may not provide accurate results. The high-degree cubature rule based CKS has developed for maintaining the high accuracy.

Fading factor based adaptive unscented two-filter smoother (AUTFS) \cite{fang2020noise} has been proposed for updating the covariance matrices. In addition, the WSLR linearized model was used to formulate the adaptive backward filter and it can run in the  forward direction. However, adaptive forward and backward filters were combined together for estimating the measurement noise matrix to obtain the smoothed solution. Sage windowing method \cite{gao2015windowing} has been developed for estimating noise statistics based on windowing approximation. This method has many advantages in the filter estimation and nonlinear systems application. In practice,  noise statistics were obtained within time window that may not accurate, eventough the filter can biased or even divergent. Random weighted estimation (RWE) method has been explored into adaptive UKF by for estimation of system noise statistics. Subsequently, the random weighting concept is adapted by adjusting random weights of each window and estimate the noise statistics online \cite{gao2015sage}. The developed RWE based adaptive UKF overcomes the limitation of the conventional UKF statistics\cite{gao2015windowing}. Adaptive random weighting based cubature Kalman flter (ARWCKF) has been developed  by adopting the concept of random weighting. This method is used to construct the random weighting estimations for estimating the noise statistics, and it adaptively adjusts the weights of cubature points to inhibit the disturbances of system noises on state estimation, leading to improve the robustness of the ARWCKS algorithm \cite{gao2019adaptively}.

\section{Cubature Kalman smoothing}
In the CKS, 2L cubature points are required to approximate the state vector mean and error covariance and followed by the spherical radial cubature criterion \cite{arasaratnam2009cubature,arasaratnam2011smoothcubature}.  

Let us consider a discrete time stochastic nonlinear dynamic system and measurement equations:

\begin{equation}\label{eqn4.1}
 {\bf x_{k}}={\bf f({x_{k-1}},u_{k-1})}+{\bf w_{k-1}}
 \end{equation}
\begin{equation}\label{eqn4.2}
 {\bf z}_{k}={\bf h({\bf x_{k}})}+{\bf v_{k}}
 \end{equation}

where, $\bf x_{k}  \in R^{n}$ is the state vector, ${\bf u_{k}}  \bf \in R^{r}$ is the input vector,  ${\bf z_{k}} \bf \in R^{m}$ is the measurement vector at time ${k}$. ${\bf f({x}_{k-1})}$ and $  {\bf h({x}_{k})}$ are the nonlinear system dynamic  and measurement functions. The process and measurement noise are assumed to be white Gaussian noise with zero mean and finite variance, represented as ${\bf w_k}=\mathbf{N}(0, {\bf Q_{k}})$ and  ${\bf v_k}=\mathbf{N}(0, {\bf R_{k}})$, respectively. The step-wise implementation of CKS algorithm is follows \cite{arasaratnam2011smoothcubature}
\subsection{Forward estimation with cubature Kalman filter}
The detailed algorithm of the CKS is given as follows.\\
Step 1: Initialize  the state estimation ${\bf \hat x}_{0}$ and error covariance matrix $ {\bf \hat P}_{0} $  as
    \begin{equation}
      \label{eqn4.3}
    \begin{cases}
    {\bf \hat x_{0}}=E[{\bf  x_{0}}]\\
     {\bf \hat P_{0}}=E[({\bf x_{0}}-{\bf \hat x_{0}})({\bf x_{0}}-{\bf \hat x_{0}})^T]
    \end{cases}
    \end{equation}
A set of 2L cubature points are given by set $ [ {({\bf \chi_k})_i} \quad { W_{i}}^{m} ] $, where ${({\bf \chi_k})_i}$ is the i th cubature point and corresponding weights are represented as 
 
 \begin{equation}
  \label{eq:t}
  \begin{aligned}
     {{\bf \chi}_i} & =\sqrt{\bf L}[1]_{i},\\        
    {\bf W_{i}}^{m} &=\frac{1}{2L}, &i=1,2..2L\\
  \end{aligned}
\end{equation}
 
 where, [1](i)  denotes its i-th column vector of the the identity matrix.
 The steps involved in the predicted (time-update) and the measurement-update of the CKS are summarised in below.
 
Prediction:\\
(1) factorise and evaluate the cubature points:

 \begin{equation}
  \label{eq:t}
  \begin{aligned}
  {\bf \hat P_{k-1}} &=S_{k-1}S_{k-1}^T,\\
  {({\bf \chi_k})_i} &=S_{k}  {{\bf \xi}_i}+ {\bf \hat x_{k}}, &i=1,2..2L\\
  \end{aligned}
\end{equation}

(2) Propagate each sigma points through the  nonlinear system  as
\begin{equation} 
{({\bf X_{k-1}})_i}= {\bf f}({{({\bf \chi_k})_i}, \bf u_{k}}),  \quad   i=1,2 ...... 2L
 \label{stateSpaceForm1}
 \end{equation}
 
(3) Evaluate the predicted  state ${\bf \hat x_{k}}^{-}$ and state error covariance ${\bf \hat P_{k}}^{-}$  based on  the  transformed sigma points  as

 \begin{equation}
  \label{eq:t}
  \begin{aligned}
{\bf \hat x_{k-1}}^{-} & =\frac{1}{2L}\sum\limits_{i=1}^{2L}{({\bf X_{k-1}})_i},\\
{\bf \hat P_{k-1}^{-}} & =\frac{1}{2L}\sum\limits_{i=1}^{2L}{({\bf X_{k-1}})_i} {({\bf X_{k-1}})_i^\top} -{\bf \hat x_{k-1}}{\bf \hat x_{k-1}}^{{-}^\top}+{\bf Q_{k-1}}\\
  \end{aligned}
\end{equation}

Measurement update:\\
(1) factorise and evaluate the new cubature points:
 \begin{equation}
  \label{eq:t}
  \begin{aligned}
  {\bf \hat P_{k-1}} &=S_{k-1}S_{k-1}^T,\\
  {({\bf Z_{k-1}})_i} &=S_{k-1}  {{\bf \xi}_i}+ {\bf \hat x_{k-1}}, &i=1,2..2L\\
  \end{aligned}
\end{equation}

(2) Propagate new cubature points through the  nonlinear system  as
\begin{equation} 
{({\bf z_{k-1}})_i}= {\bf h}({{({\bf Z_{k-1}})_i}}),  \quad   i=1,2 ...... 2L
 \label{stateSpaceForm1}
 \end{equation}
 
(3) Evaluate the predicted measurements ${\bf \hat z_{k}}^{-}$ based on the  new cubature points  as

 \begin{equation}
  \label{eq:t}
  \begin{aligned}
{\bf \hat z_{k-1}}^{-} & =\frac{1}{2L}\sum\limits_{i=1}^{2L}{({\bf Z_{k-1}})_i},\\
\end{aligned}
\end{equation}

(4) calculate the  the cross and auto covariance of state and measurement values of ${\bf P_{xz,k}} $  and  ${\bf P_{zz,k}}$ are 
\begin{equation} 
  \label{eq:t}
  \begin{aligned}
{\bf P_{xz,k-1}}&=\frac{1}{2L}\sum\limits_{i=1}^{2L} {({\bf X_k})_i}  {({\bf Z_k})_i^\top}-{\bf \hat x_{k-1}}^{-} {\bf \hat z_{k-1}}^{{-}^\top}\\
{\bf P_{zz,k-1}}&=\frac{1}{2L}\sum\limits_{i=1}^{2L} {({\bf Z_{k-1}})_i} {({\bf Z_{k-1}})_i^\top}-{\bf \hat z_{k-1}}^{-} {\bf \hat z_{k-1}}^{{-}^\top}+{\bf R_{k-1}}
\end{aligned}
 \end{equation}
 (5) evaluate the Kalman gain and the updated stat and error covariance are
\begin{equation} 
  \label{eq:t}
  \begin{aligned}
{\bf K_{k}}&={\bf P_{xz,k-1}}{\bf P_{zz,k-1}^{-1}},\\
{\bf \hat x_{k}}& ={\bf \hat x_{k-1}^{-}}+{\bf K_{k}}({\bf z_{k}}-{\bf \hat z_{k-1}^{-}})\\
{\bf \hat P_{k}} &={\bf \hat P_{k-1}}^{-}-{\bf K_{k}}{\bf P_{zz,k-1}}{\bf K_{k}}^\top.
\end{aligned}
 \end{equation}
\subsection{Backward estimation with cubature Kalman filter} 
Once forward estimation is processed, the RTS smoother is applied after the measurements. In this step, the smoother is initialized from the last time step, i.e., ${\bf \hat x_{k}^{s}}={\bf \hat x_{k}}$ and  ${\bf \hat P_{k}^{s}}={\bf \hat P_{k}}$.
The recursive process runs backwards for $ k = N-1,..., 0,$ and computes the smoother gain ${\bf K_{k}^{s}}$ , the smoothed mean and the covariance are represents ed as follows \cite{arasaratnam2011smoothcubature, psiaki2005backward}: 

Evaluate the Kalman gain and the updated stat and error covariance are
\begin{equation} 
  \label{eq:t}
  \begin{aligned}
{\bf K_{k}^{s}}&={\bf D_{k+1}}{\bf P_{zz,k+1}^{-1}},\\
{\bf \hat x_{k}^{s}}& ={\bf \hat x_{k+1}^{-}}+{\bf K_{k+1}}({\bf z_{k+1}}-{\bf \hat z_{k+1}^{-}})\\
{\bf \hat P_{k}^{s}} &={\bf \hat P_{k+1}}^{-}-{\bf K_{k+1}}{\bf P_{zz,k+1}}{\bf K_{k+1}}\top.
\end{aligned}
 \end{equation}

where,  (${\bf \upsilon_k}={\bf z_{k}}-{\bf \hat z_{k-1}^{-}}$) is the innovation sequence. ${\bf \hat P_{k}}$, is  the posterior state estimate of state. More detailed explanation of CKS can be found in  \cite{arasaratnam2011smoothcubature}
\section{Window and random weighted Adaptive Cubature RTS smoothing}
 In the CKS,  the system  and  noise models are not know exactly, hence, the filter becomes sub-optimal. In practice,  system models are nonlinear and statistical noise characteristics of the these models may vary with time. Thus, the performance of  CKS can be degraded. Consequently, the filter  becomes divergence. To address divergence issue, adaptive CKS (ACKS) has been developed based on innovation or residual for estimating the noise statistics \cite{gao2019adaptively}. In the ACKS, innovation vector is used for determining the noise statistics \cite{gao2015sage}. To improve the practicability and adaptability, windowing and random weighting estimation method has been developed. It is a promising method which can used for estimating the covariance of process and measurement noise matrices \cite{gao2015windowing, gao2019adaptively}. We utilized the same method and  develop adaptive CKS algorithm. 

\subsection{Windowing-based noise statistic estimation}

From the nonlinear system described by (3), it is obvious that
    \begin{equation}
      \label{eqn4.3}
          \begin{cases}
    E[w_{k}]=E[(x_{k}-f(x_{k}))]=0\\
     E[v_{k}]=E[(z_{k}-h(x_{k}))]=0\\
  E[w_{k}w_{k}^\top]={\bf   Q_{k-1}}= E[(x_{k}-f(x_{k}))(x_{k}-f(x_{k}))^\top]\\
    E[v_{k}v_{k}^\top]={\bf  R_{k-1}}= E[(z_{k}-h(x_{k}))(z_{k}-h(x_{k}))^\top]\\
 \end{cases}
 \end{equation}

Where, $x_{k}$ is actual state, it cannot measured directly  used in the filtering process because of the state is unobservable. Moreover, the variance of noise statistics are very small in the considering window width $N_{w}$ and uses expectation to the predicted state error   ${\hat x_{k-1}}$ and its estimated ${\hat x_{k}}$, instead of  $x_{k}$. Then, the sub-optimal estimator is defined as

    \begin{equation}
      \label{eqn4.3}
          \begin{cases}
    {\bf \hat  R_{k-j}}=\frac{1}{N_w}\sum\limits_{j=1}^{N_w}E[(x_{k}- h({\bf \hat x_{k-j}}^{-}))(x_{k}- h({\bf \hat x_{k-j}}^{-}))^\top]\\
 {\bf \hat  Q_{k-j}}=\frac{1}{N_w}\sum\limits_{j=1}^{N_w}E[(x_{k}- f({\bf \hat x_{k-j}}^{-}))(x_{k}- f({\bf \hat x_{k-j}}^{-}))^\top]
 \end{cases}
 \end{equation}

where $f({\hat x_{k-j}})$ is the posteriori mean of the estimated state through nonlinear function $f(.)$. For nonlinear CKS,  $f({\hat x_{k-j}})$ is approximated by each cubature points through nonlinear functions, that is

 \begin{equation}
  \label{eq:t}
  \begin{aligned}
 f({\bf \hat x_{k-j}}^{-}) & =\frac{1}{2L}\sum\limits_{i=1}^{2L} f({({\bf \xi_{k-j}})_i}),\\
 h({\bf \hat x_{k-1-j}}^{-}) & =\frac{1}{2L}\sum\limits_{i=1}^{2L} h({({\bf \xi_{k-1-j}})_i})\\
  \end{aligned}
\end{equation}

{\bf Theorem 1}. Suppose the system and measurement noise statistics and its variances are constants or varied with time within the window width N. Then, moving window method based  sub-optimal noise statistic estimator (19) is unbiased  for $ {\bf \hat  R_{k}}$ and $ {\bf \hat  Q_{k}}$.

    \begin{equation}
      \label{eqn4.3}
          \begin{cases}
    {\bf \hat  R_{k}}=\frac{1}{N_w}\sum\limits_{j=0}^{N_w}[{\bf \upsilon_{k-j}} {\bf \upsilon_{k-j}}^{T}-{\bf H_{k-j}} {\bf \hat P_{k-j}}  {\bf H}_{k-j}^{T}  ]\\
 {\bf \hat  Q_{k-1}}=\frac{1}{N_w}\sum\limits_{j=1}^{N_w}{\bf \hat P_{k-j}}+{\bf K_{k-j}} {\bf \upsilon_{k-j}} {\bf \upsilon_{k-j}}^{T} {\bf K}_{k-j}^{T} \\-  \sum\limits_{i=0}^{2L}{\bf W_{i}^{c}}[{({\bf X}_{k-j})_i} -{\bf \hat x_{k-j}}^{-}][{({\bf X}_{k-j})_i}-{\bf \hat x_{k-j}}^{-}]^{T}
 \end{cases}
 \end{equation}

 \begin{proof}
The innovation sequence is described as
 \begin{equation}
 {\bf \upsilon_{k}}={\bf z_{k}}-{\bf \hat z_{k}^{-}}
  \label{stateSpaceForm1}
  \end{equation}
and substituting the measurement equation in (2) into (21), we have

  \begin{equation}
 {\bf \upsilon_{k}}={\bf h(.)}({\bf {x}_{k}}-{\bf \hat {x}^{-}_{k}})+{\bf v_k}
  \label{stateSpaceForm1}
  \end{equation}

By taking the expectation of the innovation sequence is  

    \begin{equation}
      \label{eqn4.3}
          \begin{cases}
    E[{\bf \upsilon_{k}}]=E[{\bf z_{k}}-{\bf \hat z_{k}^{-}}]=0\\
  E[{\bf \upsilon_{k}}{\bf \upsilon_{k}}^\top]= E[({\bf z_{k}}-{\bf \hat z_{k}^{-}})({\bf z_{k}}-{\bf \hat z_{k}^{-}})^\top]={\bf P_{zz,k}}\\
 \end{cases}
 \end{equation} 

 The window width is ${N_w}$ and there are ${N_w}$ measurements within $t_{k-N}$ to $  t_{k}$. In this stage, the noise statistics is very small variation  within a window width. Define the predicted and estimated state error are 
 \begin{equation}
  \label{eq:t}
  \begin{aligned}
{\bf \Delta \hat {x}^{-}_{k-j}}&={\bf \hat {x}_{k-j}}-{\bf \hat {x}^{-}_{k-1-j}},\\
&= {\bf K_{k-j}}({\bf z_{k-j}}-{\bf \hat z_{k-1-j}^{-}})\\
&= {\bf K_{k-j}}{\bf \upsilon_{k-j}}
  \end{aligned}, j=1,2,...N
\end{equation}  
From equations (21) and (22), we can apply the expectation 
 \begin{equation}
  \label{eq:t}
  \begin{aligned}
E[{\bf w_{k-1}}]&=\frac{1}{N}\sum\limits_{i=1}^{N} E[ {\bf \hat {x}^{-}_{k}} - \sum\limits_{i=1}^{N}f({({\bf \xi_{k-j}})_i})]\\
&= \frac{1}{N}\sum\limits_{i=1}^{N} E[ {\bf \hat {x}^{-}_{k-j}} - {\bf \hat {x}^{-}_{k-1-j}}]\\
&=\frac{1}{N}\sum\limits_{i=1}^{N} E[ {\bf \hat {x}^{-}_{k-j}} - {\bf \hat {x}^{-}_{k-1-j}}] \\
&=\frac{1}{N}\sum\limits_{i=1}^{N} [ {\bf K_{k-j}}{\bf \upsilon_{k-j}}] \\
&=0
  \end{aligned}
\end{equation}

and, in the above equation represents the mean and  covariance of the predicted state after transferred by nonlinear function. The sub-optimal unbiased  estimator under the measurement noise vector can be obtained as follows;
 \begin{equation}
  \label{eq:t}
  \begin{aligned}
E[{\bf v_{k}}]&=\frac{1}{N}\sum\limits_{i=1}^{N} E[ {\bf {Z}_{k}} - \sum\limits_{i=1}^{N}h({({\bf \xi_{k-j}})_i})]\\
&= \frac{1}{N}\sum\limits_{i=1}^{N} E[ {\bf \hat {x}^{-}_{k-j}} - {\bf \hat {x}^{-}_{k-1-j}}]\\
&=\frac{1}{N}\sum\limits_{i=1}^{N} E[ {\bf \hat {x}^{-}_{k-j}} - {\bf \hat {x}^{-}_{k-1-j}}] \\
&=\frac{1}{N}\sum\limits_{i=1}^{N} [ {\bf K_{k-j}}{\bf \upsilon_{k-j}}] \\
&=0
  \end{aligned}
\end{equation}

 \begin{equation}
  \label{eq:t}
  \begin{aligned}
E[{\bf \hat  Q_{k-1}}]&=\frac{1}{N}\sum\limits_{i=1}^{N} E[ {\bf \hat {x}_{k-j}} - f({\bf \hat {x}_{k-1-j}^{-}} ) ]\\
&= \frac{1}{N}\sum\limits_{i=1}^{N} E[ {\bf \hat {x}^{-}_{k-j}} - {\bf \hat {x}^{-}_{k-1-j}}]\\
&=\frac{1}{N}\sum\limits_{i=1}^{N} [ {\bf K_{k-j}} {\bf \upsilon_{k-j}} {\bf \upsilon_{k-j}^\top} {\bf K_{k-j}^\top} ] \\
&=\frac{1}{N}\sum\limits_{i=1}^{N} [ {\bf K_{k-j}} {\bf \hat P_{zz, k-j}} {\bf K_{k-j}^\top} ] \\
&=\frac{1}{N}\sum\limits_{i=1}^{n} [ {\bf \hat P_{k-1-j}^{-}}-{\bf \hat P_{k-j}}  ] \\
 &=\frac{1}{N} \sum\limits_{i=1}^{N} [\frac{1}{2L}\sum\limits_{i=1}^{2L}{({\bf X_{k-1}})_i} {({\bf X_{k-1}})_i^\top} -{\bf \hat x_{k-1}}{\bf \hat x_{k-1}}^{{-}^\top} \\
 &  +{\bf Q_{k-1}}-{\bf \hat P_{k-j}} ] \\
 &=\frac{1}{N} \sum\limits_{i=1}^{N} [\frac{1}{2L}\sum\limits_{i=1}^{2L}{({\bf X_{k-1}})_i} {({\bf X_{k-1}})_i^\top} -{\bf \hat x_{k-1}}{\bf \hat x_{k-1}}^{{-}^\top}\\ &-{\bf \hat P_{k-j}} ] +{\bf Q_{k-1}}  \\ 
 &={\bf Q_{k-1}}  \\
 \end{aligned}
\end{equation}

Thus, the unbiased estimation for $Q_{k-1}$ can be represented as

 \begin{gather}
  {\bf \hat Q_{k-1}} =\frac{1}{N} \sum\limits_{i=1}^{N} [{\bf \hat P_{k-j}} + {\bf K_{k-j}} {\bf \upsilon_{k-j}} {\bf \upsilon_{k-j}^\top} {\bf K_{k-j}^\top}  \\ \notag -\frac{1}{2L}\sum\limits_{i=1}^{2L}{({\bf X_{k-1}})_i} {({\bf X_{k-1}})_i^\top} -{\bf \hat x_{k-1}}{\bf \hat x_{k-1}}^{{-}^\top}  
 \end{gather}

From equation (22), we can apply the expectation 
 \begin{equation}
  \label{eq:t}
  \begin{aligned}
E[{\bf \hat R_{k}}]&=\frac{1}{N}\sum\limits_{j=1}^{N} E[ {\bf {Z}_{k-j}} - \sum\limits_{i=1}^{2L}h({({\bf \xi_{k-j}})_i})]\\
&= \frac{1}{N}\sum\limits_{j=1}^{N} E[ {\bf \upsilon_{k-j}} {\bf \upsilon_{k-j}^\top}]\\
&= \frac{1}{N}\sum\limits_{j=0}^{N-1}{\bf P_{zz, k-j}}  \\
&= \frac{1}{N}\sum\limits_{j=1}^{N} [\sum\limits_{i=1}^{2L} {({\bf Z_{k-1}})_i} {({\bf Z_{k-1}})_i^\top}-{\bf \hat z_{k-1}}^{-} {\bf \hat z_{k-1}}^{{-}^\top}+{\bf R_{k-1}}]\\
&= \frac{1}{N}\sum\limits_{j=1}^{N} \sum\limits_{i=1}^{2L} {({\bf Z_{k-1}})_i} {({\bf Z_{k-1}})_i^\top}-{\bf \hat z_{k-1}}^{-} {\bf \hat z_{k-1}}^{{-}^\top}+{\bf R_{k-1}}\\
  \end{aligned}
\end{equation}

Thus, the unbiased estimation for $R_{k-1}$ can be represented as

 \begin{gather}
 {\bf \hat R_{k-1}} =\frac{1}{N} \sum\limits_{i=1}^{N} [{\bf \upsilon_{k-j}} {\bf \upsilon_{k-j}^\top} -\sum\limits_{i=1}^{2L} {({\bf Z_{k-1}})_i} {({\bf Z_{k-1}})_i^\top}-\\{\bf \hat z_{k-1}}^{-} {\bf \hat z_{k-1}}^{{-}^\top} ]  
 \end{gather}
where ${\bf \upsilon_{k-j}}$ is the innovation vector. The proof the unbiasedness of the estimator (25) can be easily proved. The remark of a nonlinear Gaussian system, the posterior mean and covariance approximation of the state through cubature propagation,  which means the residual sequence, i.e., ($E[{\bf \upsilon_{k-j}}]=0$),  of filter is an approximate Gaussian white noise sequence. \cite{gao2019adaptively}. 

\end{proof}

{\bf B. Noise Statistic Estimator Based on Moving Windowing and Random Weighting Methods:}

In theorem 1, we presented a simple  moving window based noise statistic estimator \cite{gao2015windowing}. However, system and measurement noise statistics are changing rapidly in practice, thus, the moving window based estimator is used to compute the noise statistics based on true condition of the system’s noises at the current epoch. Moreover, the residual weights are the  same  and also too much historical information are required in this case.  To address this issue, the to combined the windowing and random weighting method is used to adjust the weights on the residuals, which will shows the performance improvement of the filter.

{\bf Theorem 2}. To utilize the theorem 1, noise statistic estimator of the moving window and random weighting methods can be described as

\begin{equation}
\begin{split}
  {\bf \hat  Q}_{k-1}^{*}=\sum\limits_{j=1}^{N}{\bf \lambda_{j}}[{\bf \hat P_{k-j}}+{\bf K_{k-j}} {\bf \varepsilon_{k-j}} {\bf \varepsilon_{k-j}}^{T} {\bf K}_{k-j}^{T}\\ -  \sum\limits_{i=0}^{2L}{\bf W_{i}^{c}}[{({\bf X}_{k-j})_i} -{\bf \hat x_{k-j}}^{-}][{({\bf X}_{k-j})_i}-{\bf \hat x_{k-j}}^{-}]^{T}]
  \label{stateSpaceForm1}
  \end{split}
 \end{equation}

\begin{equation} 
{\bf \hat  R}_{k-1}^{*}=\sum\limits_{j=0}^{N} {\bf \lambda_{j}}[{\bf \upsilon_{k-j}} {\bf \upsilon_{k-j}}^{T}-{\bf H_{k-j}} {\bf \hat P_{k-j}}  {\bf H}_{k-j}^{T}  ] 
 \label{stateSpaceForm1}
 \end{equation}.

where,  $\ [\bf \lambda_{1}, \bf \lambda_{2},.......... \bf\lambda_{n} \ ]$ is are the random vector and it follows Dirichlet distribution,  D(1, 1, ..., 1),
i.e., $ \sum_{i=1}^{n}{\bf \lambda_{i}}=1$.   Then, evaluate the joint density function of $\ [\bf \lambda_{1},\bf \lambda_{2},.......... \bf \lambda_{n} \ ]$ is $ \bf f(\lambda_{1},\lambda_{2},..........\lambda_{n})={\bf \tau_{n} }$. where $(\bf  \lambda_{1}, \bf \lambda_{2},.......... \bf \lambda_{n} ) \bf \epsilon  \bf  D_{n}$ and $ \bf D_{n-1}=\{ [\bf \lambda_{1}, \bf \lambda_{2},.......... \bf \lambda_{n} : \bf \lambda_{k} > 0 \ (k=1,2,...n-1),\  \sum_{i=1}^{n-1}{\bf \lambda_{k}}< 1\} $. If the noise statistics are constants or the variations of them are
very small in the window width N,  satisfying (18), then the moving window and random weighting methods of the noise statistic estimator is unbiased.
\begin{proof}
By taking the random weighting estimation principle, It can be easily obtained noise statistics, Here,  we prove that an estimator satisfies the unbiasedness.

Taking the expectation of equation (\ref{eq:31}) yields

 \begin{equation}
  \label{eq:31}
  \begin{aligned}
 E[{\bf \hat  Q}_{k-1}^{*}]&=\frac{1}{N}\sum\limits_{i=1}^{N} {\lambda_{j}}  E[ {\bf \hat {x}_{k-j}} - f({\bf \hat {x}_{k-1-j}^{-}} ) ]\\
&= \frac{1}{N}\sum\limits_{i=1}^{N} {\lambda_{j}} E[ {\bf \hat {x}^{-}_{k-j}} - {\bf \hat {x}^{-}_{k-1-j}}]\\
&=\frac{1}{N}\sum\limits_{i=1}^{N} {\lambda_{j}} [ {\bf K_{k-j}} {\bf \upsilon_{k-j}} {\bf \upsilon_{k-j}^\top} {\bf K_{k-j}^\top} ] \\
&=\frac{1}{N}\sum\limits_{i=1}^{N} {\lambda_{j}} [ {\bf K_{k-j}} {\bf \hat P_{zz, k-j}} {\bf K_{k-j}^\top} ] \\
&=\frac{1}{N}\sum\limits_{i=1}^{n} {\lambda_{j}} [ {\bf \hat P_{k-1-j}^{-}}-{\bf \hat P_{k-j}}  ] \\
 &=\frac{1}{N} \sum\limits_{i=1}^{N} {\lambda_{j}}  [\frac{1}{2L}\sum\limits_{i=1}^{2L}{({\bf X_{k-1}})_i} {({\bf X_{k-1}})_i^\top} \\ -{\bf \hat x_{k-1}}{\bf \hat x_{k-1}}^{{-}^\top} \\
 &  +{\bf Q_{k-1}}-{\bf \hat P_{k-j}} ] \\
 &=\frac{1}{N} \sum\limits_{i=1}^{N} {\lambda_{j}}  [\frac{1}{2L}\sum\limits_{i=1}^{2L}{({\bf X_{k-1}})_i} {({\bf X_{k-1}})_i^\top} \\-{\bf \hat x_{k-1}}{\bf \hat x_{k-1}}^{{-}^\top} &-{\bf \hat P_{k-j}} ] +{\bf Q_{k-1}}  \\ 
 &=\frac{1}{N} \sum\limits_{i=1}^{N}{\lambda_{j}} {\bf Q_{k-1}}  \\
  &={\bf Q_{k-1}}  \\
 \end{aligned}
\end{equation}

 and

 \begin{equation}
  \label{eq:t}
  \begin{aligned}
E[{\bf \hat  R}_{k-1}^{*}]&=\frac{1}{N}\sum\limits_{j=1}^{N} {\lambda_{j}}  E[ {\bf {Z}_{k-j}} - \sum\limits_{i=1}^{2L}h({({\bf \xi_{k-j}})_i})]\\
&= \frac{1}{N}\sum\limits_{j=1}^{N} {\lambda_{j}}  E[ {\bf \upsilon_{k-j}} {\bf \upsilon_{k-j}^\top}]\\
&= \frac{1}{N}\sum\limits_{j=0}^{N-1}{\bf P_{zz, k-j}}  \\
&= \frac{1}{N}\sum\limits_{j=1}^{N} {\lambda_{j}} [\sum\limits_{i=1}^{2L} {({\bf Z_{k-1}})_i} {({\bf Z_{k-1}})_i^\top}-{\bf \hat z_{k-1}}^{-} {\bf \hat z_{k-1}}^{{-}^\top} + \\ {\bf R_{k-1}}]\\
&= \frac{1}{N}\sum\limits_{j=1}^{N} {\lambda_{j}}  [ \sum\limits_{i=1}^{2L} {({\bf Z_{k-1}})_i} {({\bf Z_{k-1}})_i^\top}-{\bf \hat z_{k-1}}^{-} {\bf \hat z_{k-1}}^{{-}^\top}+ \\{\bf R_{k-1}}]\\
&= \frac{1}{N}\sum\limits_{j=1}^{N}  {\lambda_{j}} {\bf R_{k-1}}\\
&= {\bf R_{k-1}}\\
  \end{aligned}
\end{equation}

The equations (33) and (34) imply that an estimator is unbiased.
\end{proof}
\subsection{ Determination of Random Weighting Factors}
Suppose the  predicted (${\hat x_{k-1}}$) and estimated of the state (${\hat x_{k-j/k-1-j}}$) at epoch ${k-j}$ (j=1,2,..n), respectively. The residual vector (${\Delta x_{k-1}}$) of the prediction of the state is assumed as  

\begin{equation} 
{\Delta x_{k-1}}={\hat x_{k-1}}- {\hat x_{k-j/k-1-j}}
 \label{stateSpaceForm1}
 \end{equation}.
The residual vector of the measurement is expressed as

\begin{equation} 
{\Delta z_{k-1}}={\hat z_{k-1}}- {\hat z_{k-j/k-1-j}}
 \label{stateSpaceForm1}
 \end{equation}.

If noise statistics of the system is changed, then  the predicted state, ${\hat x_{k-j/k-1-j}}$,  will decrease, which means leading the predicted state to be biased. As a result, the magnitude of the residual vector of predicted state  ${\Delta x_{k-1}}$ill can increase. Similarly, when the  measurement noise statistics are changed, the residual vector of  measurement  ${\Delta z_{k-1}}$  will be biased and also  magnitude of residual vector of measurements will increase. Hence, random weighting factors are required to 
to capture the changes of noise of the system and also which can satisfy as

\begin{equation} 
{\lambda_{j}} \propto \|{\Delta x_{k-j}}\| \|{\Delta z_{k-j}}\|
 \label{stateSpaceForm1}
 \end{equation}.

where $\|{\Delta x_{k-1}}\|= \sqrt{{\Delta x_{k-1}} {\Delta x_{k-1}}^{T}}$,  $\|{\Delta z_{k-1}}\|= \sqrt{{\Delta z_{k-1}} {\Delta z_{k-1}}^{T}}$ and the symbol $\propto$ indicates the proportional operation.

In the above Equation, weighting factor is proportional to the $\|{\Delta x_{k-1}}\|$ and  $\|{\Delta z_{k-1}}\|$. If  weighting factor is increased whereas an residual error values can be increased because of proportional. In general, covariance matching method is an effective way to detect and  eliminate the disturbance and abnormality  in measurements by  adjusting their weights to the filter,  getting the solution through the following inequality.

 \begin{equation}
  \label{eq:t}
 \begin{split}
[{\bf \upsilon_{k-j}^\top} {\bf \upsilon_{k-j}}] \leq  S\quad tr[ E[{\bf \upsilon_{k-j}} {\bf \upsilon_{k-j}^\top}] ], (j=1,2,3...,N)\\
= S \quad tr[\sum\limits_{i=1}^{2L} {({\bf Z_{k-1}})_i} {({\bf Z_{k-1}})_i^\top}- {\bf \hat z_{k-1}}^{-} {\bf \hat z_{k-1}}^{{-}^\top} +  {\bf \hat R_{k-1}} ]
 \end{split}
\end{equation}

where ${\bf \upsilon_{k-j}}$ is the innovation vector and S is an adjustable factor satisfying S $\geq$ 1

In this paper, because ${\bf \hat R_{k-1}}$ is unknown, by replacing ${\bf \hat R_{k-1}}$ with its estimate ${\bf \hat R_{k-1}^{*}}$ , (38) can be rewritten as
 \begin{equation}
  \label{eq:37}
 \begin{split}
[{\bf \upsilon_{k-j}^\top} {\bf \upsilon_{k-j}}] \leq  S \quad tr[\sum\limits_{i=1}^{2L} {({\bf Z_{k-1}})_i} {({\bf Z_{k-1}})_i^\top}- {\bf \hat z_{k-1}}^{-} {\bf \hat z_{k-1}}^{{-}^\top}  \\+  {\bf \hat R_{k-1}^{*}} ]
 \end{split}
\end{equation}

If the equation (\ref{eq:37}) is not satisfied, the  weight on the $k-j$ th residual in measurements, that shows  abnormal measurement and ought to be small. Thus, the random weighting factors are also required to satisfy

\begin{equation}
\begin{split}
{\lambda_{j}} \propto \frac{ S \quad tr[\sum\limits_{i=1}^{2L} {({\bf Z_{k-1}})_i} {({\bf Z_{k-1}})_i^\top}- {\bf \hat z_{k-1}}^{-} {\bf \hat z_{k-1}}^{{-}^\top} +  {\bf \hat R_{k-1}^{*}} ]}{[{\bf \upsilon_{k-j}^\top} {\bf \upsilon_{k-j}}]}\\
=\Delta S(j)
 \label{stateSpaceForm1}
 \end{split}
 \end{equation}.

Therefore, weights on $k-j$th residual is evaluated and corresponding the random weighting factor (${\lambda_{j}}$ ) can be determined as follows.
\begin{equation} 
{\omega_{j}} = \|{\Delta x_{k-j}}\| \|{\Delta z_{k-j}}\| \Delta S(j) \quad (j=1,2,..N)
 \label{stateSpaceForm1}
 \end{equation}.
Normalizing the  ${\omega_{j}}$(j=1,2,..n) of the random weighting
factors are obtained as

\begin{equation} 
{\lambda_{j}} = \frac{{\omega_{j}}}{\sum\limits_{j=1}^{N} {\bf \omega_{j}}}
 \label{stateSpaceForm1}
 \end{equation}.
 
 where ${\lambda_{1}}, {\lambda_{2}},...{\lambda_{N}}$ obeys Dirichlet distribution, D(1,1,1...1).
The innovation vectoris used to estimated the noise statistics and followed by the windowing and random weighted estimation. It enables adaptively adjust the weights on each residual or innovation vector to improve the filter accuracy. However, the process noise and measurement noise on state estimation can also improve the reliability of the filter.

\subsection{Forward estimation with WRWACRTS  algorithm}
In the proposed method,  predicted and measurement update phase are  updated adaptively by random weighted factors. The estimated state and its error covariance equations are involved in the measurement updated equations. The updated predicted equations are  

 \begin{equation}
  \label{eq:t}
  \begin{aligned}
{\bf \hat x_{k-1}}^{-} & =\frac{1}{2L}\sum\limits_{i=1}^{2L}{({\bf X_{k-1}})_i},\\
{\bf \hat P_{k-1}}^{-} & =\frac{1}{2L}\sum\limits_{i=1}^{2L}{({\bf X_{k-1}})_i} {({\bf X_{k-1}})_i^\top} -{\bf \hat x_{k-1}}{\bf \hat x_{k-1}}^{{-}^\top}+{\bf \hat Q}_{k}^{*}\\
  \end{aligned}
\end{equation}


In this step, adaptive Kalman gain ${\bf K_k}$ is updated by

\begin{equation} 
{\bf K_{k}}={\bf P_{xz,k}}{\bf P_{zz,k}^{-1}}
 \label{stateSpaceForm1}
 \end{equation}

Where  ${\bf P_{xz,k}} $ is the cross covariance of state and measurement values.
\begin{equation} 
{\bf P_{xz,k}}=\sum\limits_{i=0}^{2L}{\bf W_{i}^{c}}[ {({\bf X_k})_i}-{\bf \hat x_{k}}^{-}][ {({\bf Z_k})_i}-{\bf \hat z_{k}}^{-}]^{T}
 \label{stateSpaceForm1}
 \end{equation}
and ${\bf P_{zz,k}}$ is the auto covariance of innovation sequence evaluated as
\begin{equation} 
{\bf P_{zz,k}}=\sum\limits_{i=0}^{2L}{\bf W_{i}^{c}}[{({\bf Z_k})_i}-{\bf \hat z_{k}}^{-}] [ {({\bf Z_k})_i}-{\bf \hat z_{k}}^{-}]^{T}+{\bf R}_{k}^{*}
 \label{stateSpaceForm1}
 \end{equation}




\subsection{Backward estimation with WRWACRTS  algorithm} 
Once forward estimation is processed, the RTS smoother is applied after the measurements. In this step, recursive process runs backwards for $ k$ = ${N-1}$,..., 0 and computes the smoother gain ${\bf K_{k}^{s}}$, the smoothed mean and the covariance are represented as follows. The Kalman gain, the updated state and error covariance are evaluated as
\begin{equation} 
  \label{eq:t}
  \begin{aligned}
{\bf K_{k}^{s}}&={\bf D_{k+1}}{\bf P_{zz,k+1}^{-1}},\\
{\bf \hat x_{k}^{s}}& ={\bf \hat x_{k+1}^{-}}+{\bf K_{k+1}}({\bf z_{k+1}}-{\bf \hat z_{k+1}^{-}})\\
{\bf \hat P_{k}^{s}} &={\bf \hat P_{k+1}}^{-}-{\bf K_{k+1}}{\bf P_{zz,k+1}}{\bf K_{k+1}}\top.
\end{aligned}
 \end{equation}

where,  (${\bf \upsilon_k}={\bf z_{k}}-{\bf \hat z_{k-1}^{-}}$) is the innovation sequence. ${\bf \hat P_{k}}$, is  the posterior state estimate of state. More detailed explanation of CKF can be found in  \cite{arasaratnam2009cubature}

\section{Numerical Simulation}

In this section, to show the performance of the proposed adaptive algorithm is demonstrated by a nonlinear target tracking example has been considered. It is a benchmark
problem that has been used to test the effectiveness of different nonlinear adaptive filters\cite{arasaratnam2011smoothcubature}. The nonlinear state and measurement model of tracking example can be expressed as follows \cite{wan2000unscented}:

\begin{equation}
      \label{eqn4.3}
    \begin{cases}
    {x_{1}}(k+1)&={x_{1}}(k)+T_s{x_{3}}(k)+{ w_{1,k}}\\

    {x_{2}}(k+1)&={x_{2}}(k)+T_s( -{k_{x}} { x_{3}}^2)(k)+{ w_{2,k}}\\
     {x_{3}}(k+1)&={x_{3}}(k)+T_s{ x_{4}}(k)+{ w_{3,k}}\\   
    
    { x_{4}}(k+1)&={ x_{4}}(k)+T_s( { k_{y}} { x_{3}}^2-g)(k)+{ w_{4,k}}\\
     \end{cases}
    \end{equation}
    
The model of the target motion state is as follows:
\begin{equation}\label{eqn4.1}
 {\bf x_{k}}={\bf F{x_{k-1}}}+{\bf w_{k}}
 \end{equation}

The target motion state vector is a linear cases, F  is linear state transition matrix is designed based on equation (48), which describes the motion of state. Where, the state, ${x_{k}}=[{x_{1,k}} \quad {x_{2,k}} \quad  {x_{3,k}} \quad {x_{4,k}} ]^T$ are the vehicle position and velocity in $x-y$ plane and its constant coefficients. ${w_{k}}= [ { w_{1,k}} \quad  {w_{2,k}} \quad {w_{3,k}} \quad {w_{4,k}}]$ are the process noises. $T_s$ is the step size is set to  0.1 s. The $g=9.8 m/s^2$ . The ${w_{k}}$  is the assumed to be white Gaussian process noise with zero mean and covariance. 


The nonlinear measurement model is given by 
\begin{equation}
      \label{eqn4.3}
    \begin{cases}
      { z_{1}}(k)&= \sqrt{( x_{1}({k})^2+( x_{3}(k)^2} +   {v_{1}}(k)  \\  { z_{2}}(k)&= {\tan^{-1}( \frac{{ x_{3}}(k)}{ { x_{1}}(k)}) } + { v_{2}(k)} \\     \end{cases}
    \end{equation}

where, $z_{k}=[ z_{1,k} \quad z_{2,k}]$ are the range and angle measurements. tan is the four-quadrant inverse tangent function; $v_{k}=[ v_{1,k} \quad v_{2,k}]$ are the measurement noises and is assumed to be the white Gaussian measurement noise with zero mean and its covariance. Equations (49) and (50) are the state  and measurement equations, state is a linear, whereas the measurement equation is a non-linear. Since the measurement equation is a non-linear, then, the  problem of target tracking is a non-linear system in this study \cite{fang2020noise}. The process noise covariance matrix is $Q_{k}$ is represents as

 ${Q_{k}}$=\[
  \begin{bmatrix}
    2\times10^{-1} & 0 & \dots & 0 \\
    0 & 2\times10^{-1} & \dots & 0 \\
    0 & 0 & 2\times10^{-1} & 0 \\
    0 & 0 & \dots & 2\times10^{-1}
  \end{bmatrix}
\]

The initial measurement noise covariance matrix,  $R_{k} = diag([100 \quad 3\times 10^{-3}])$is selected. The initial estimate ${\bf \hat x_{k}}$ is generated randomly from
the normal distribution ${\mathbf{N}}({\bf \hat x_{0}}; {\bf x_{0}}, {\hat P_{0}})$. Where, ${\bf x_{0}}$ is the actual initial sate $x_{0}$  = $[0 m,\quad  0 m/s \quad 0 m  \quad 0 m]^\top $  and initial error covariance,  ${\hat P_{0}}$ = diag($[100 m, \quad  100 m/s, \quad 100 m, \quad 100 m/s]$). The proposed algorithm is applied to a target tracking example in comparison with the standard CKS and ACKS. The true trajectory and estimated target trajectory is shown on Figure 1. In this analysis,  We selected an optimal window width is N=15 samples.
 
\begin{figure}[!ht]
\centering
\includegraphics[scale=0.55]{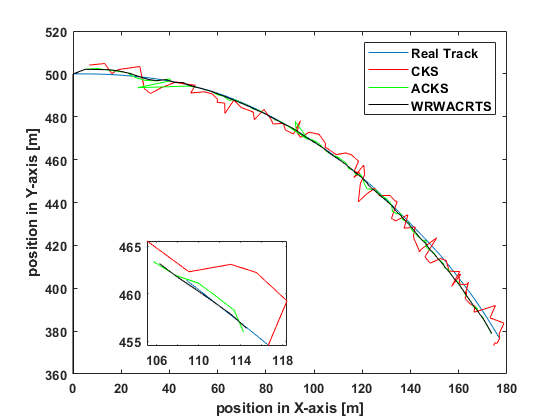}
 \caption{Trajectory of the maneuvering target. CKS: Cubature Kalman smoothing; ACKS: Adaptive cubature Kalman smoothing; WRWACRTS: Windowing and random weighted estimation based adaptive cubature RT smoothing}
\label{fig:2}
\end{figure}
\begin{figure}[!ht]
\centering
\includegraphics[scale=0.55]{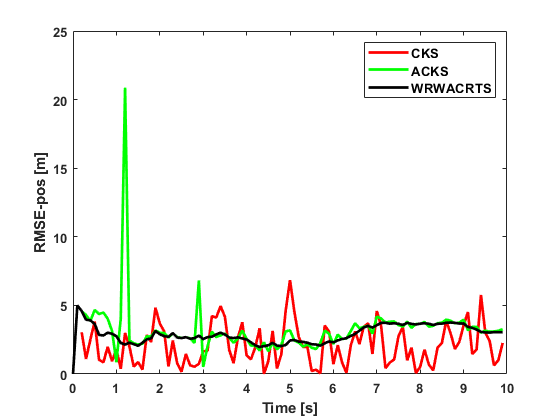}
 \caption{Position RMSE of proposed algorithms.}
\label{fig:2}
\end{figure}

\begin{figure}[!ht]
\centering
\includegraphics[scale=0.55]{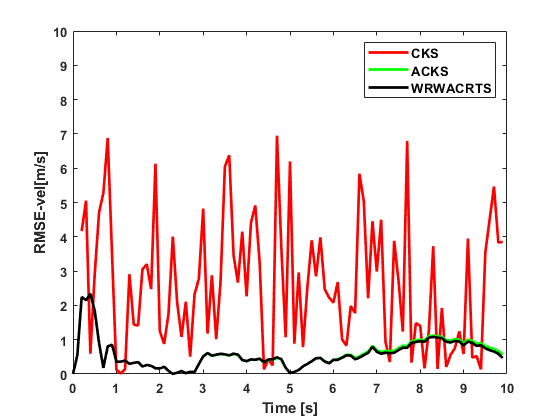}
 \caption{Speed RMSE of proposed algorithms.}
\label{fig:2}
\end{figure}

The root mean square error (RMSE) is used  to compare the position and velocity estimation of proposed algorithms. The RMSE formulation  is given by

\begin{equation}
      \label{eqn4.3}
      {\bf RMSE}= \sqrt{\frac{\sum\limits_{i=1}^{N_{sim}} (\bf x_{i}-\bf \hat x_{i})^2}{N_{sim}} }  
      \end{equation}
      
where $N_{sim}$ = 100 is the number of  runs.  $\bf x_{k}$ and $\bf \hat x_{k}$ are the actual and estimated values of the position, respectively. The similar formula is used for  RMSE velocity calculation.  
      
Fig. 1 illustrates the actual and estimated target trajectory performance with the CRTS and the proposed algorithm. It can be seen that the estimation accuracy of the  conventional CKS is not exactly follow the true  state of $x_{1,k}$ and filter becomes diverge because of inaccurate of process and measurement noise covariance matrices. Moreover, the adaptive CRTS and proposed algorithms, the noise statistics are updated adaptively. It shows that the ACKS and WRWACRTS can track the true state $x_{1,k}$ exactly and  filter accuracy is improved. Fig. 2  and Fig.3 illustrates position and speed error of the filters for target tracking examples. In CKS, the process noise covariance is constant value. From Fig.2,    conventional CKS leads to large estimation error in position.. However, WRWACKS algorithm is  the good performance than the CKS and ACKS and its outstanding merits in target tracking. The  RMSE values  of  proposed algorithms are tabulated in Table I.

\begin{table}[ht]
\caption{ Average RMSE values  of the proposed algorithm }
\label{tab:1}  
\begin{tabular}{lllll}
\hline\noalign{\smallskip}
 & Position RMSE [m] &  Speed RMSE[m/s]\\
\noalign{\smallskip}\hline\noalign{\smallskip}
 CKS &  1.758 &    0.604\\
ACKS & 1.391 &  0.507\\
WRWACRTS & 0.550 & 0.389\\
\noalign{\smallskip}\hline
\end{tabular}
\end{table}

This demonstrates that the proposed algorithm  can track the true trajectory and also updated the noise statistic online.  The performance of the proposed algorithm is improved in terms minimum RMSE as compared to conventional CKS and adaptive CKS.

\section{Conclusion}

In this paper, we combine windowing and random weighting
concepts and applied into cubature Kalman smoothing to developed a novel adaptive cubature Kalman smoothing (WRWACRTS) algorithm. Windowing theory is firstly addressed and then to extend the random windowing concept  to adaptive cubature Kalman smoothing. secondly, combine windowing and random weighted theories are utilized and to adjust random
weights dynamically based on historical residuals for estimation of system and measurement noise statistics. However, the proposed WRWACRTS method overcomes the problem of the conventional CKS and adaptive CKS in requiring precise knowledge on statistical
characteristics of system noise. Moreover, it can provide higher filtering accuracy over the CKS and ACKS algorithm. The algorithm can also requires small computation time, sufficient to achieve the good performance in simulation example. Based on the RMSE values are observed in simulation result reveal that the proposed algorithms outperforms the other algorithms.

\appendix{Convergence analysis of proposed algorithm}
In this section,  we used Lyapunov stability for analyzing the convergence of the proposed algorithm and  following proposition \cite{zarei2014convergence, fang2020noise}.
Consider the WRWACRTS described in previous section. The convergence of the WRWACRTS is ensured if the following holds
 \begin{equation}
 \begin{split}
(1 -\frac{2}{\alpha_k}) {\bf \hat R_{k}^{-1}} + {\frac{1}{\bf \alpha_{k}^{2}}} {\bf \hat R_{k}^{-1}}{ \bf H} {\bf \hat P_{k}} { \bf H^\top} {\bf \hat R_{k}^{-1}} \le 0\\
{\bf \alpha_{k}} {\bf F_{k}}^{\top} {\bf \zeta_{k}}({\bf \hat P_{k}}^{-})^{-1} {\bf \zeta_{k}} {\bf F_{k}}-{\bf \hat P_{k-1}^{-1}} \le 0,
\end{split}
\label{condition}
\end{equation}
  
 where,
 \begin{eqnarray*}
{\bf \zeta_{k}}&=&{\mathrm {diag}} \{ {\bf \zeta_{1k}}, \dots , {\bf \zeta_{mk}}  \}.
  \end{eqnarray*}
  \begin{eqnarray*}
{\bf \hat P_{k}}^{-}&=&\frac{1}{2L}\sum\limits_{i=1}^{2L}{({\bf X_{k}})_i} {({\bf X_{k}})_i^\top} -{\bf \hat x_{k}}{\bf \hat x_{k}}^{{-}^\top}+{\bf Q_{k}}.
  \end{eqnarray*} 

  \begin{proof}
The state estimation and prediction error vectors are defined as
\begin{eqnarray*}
 { \bf  \tilde x_{k}}&=&{\bf {  x_{k}}}-{ \bf \hat x_{k}} \\
 { \bf  \tilde x_{k}}^{-}&=&{\bf {  x_{k}}}-{ \bf \hat x_{k}^{-}}
\label{egn:error}
  \end{eqnarray*}
 Consider the candidate   Lyapunov function
\begin{equation}
 { \bf V_{k} }= {\bf \tilde x_{k}}^{\top} {\bf P_{k}^{-1}}{\bf \tilde x_{k}}.
\label{Lyap}
  \end{equation}
  The  objective is to derive conditions under which the sequence $ \{{ \bf V_{k} }\}_{k=1}^{\infty}$  is decreasing ($ { \bf V_{k+1} }- { \bf V_{k} }\le 0$) along the trajectories of the cubature Kalman filter
 \begin{eqnarray}
 \begin{array}{lcl}
{\bf x_{k}}&=& f ({ \bf  x_{k-1}})+ {\bf w_{k-1}}\\
{\bf z_{k}}&=& h({ \bf  x_{k}})+ {\bf v_{k}}
\end{array}
\label{egn:state_space}
  \end{eqnarray}
  
 For convenience, we use the presented approach in [24, 25], to simplify the error expression 

  \begin{eqnarray}
 \begin{array}{lcl}
{\bf F_{k}}&=& [ \frac{\partial f ({ \bf  x_{k-1}})}{\partial x}]_{{ \bf  x_{k}}= { \bf \hat x_{k}}}\\
{\bf H_{k}}&=& [\frac{\partial h ({ \bf  x_{k-1}})}{\partial x}]_{{ \bf  x_{k}}= { \bf \hat x_{k}}}
\end{array}
\label{egn:state_space}
  \end{eqnarray} 

where, $ {\bf w_{k-1}}$  and $ {\bf v_{k}}$ are the white noise sequence with  zero mean. The innovation error, defined as   the difference between measured output and the predicted measurement is
 \begin{eqnarray}
 \begin{array}{lcl}
 {\bf e_{k}}&=& {\bf z_{k}}- { \bf \hat z_{k}^{-}}\\
& =&  {\bf H_{k}} ({\bf  x_{k}}+{\bf v_{k}})-{\bf H_{k}}{ \bf \hat x_{k}^{-}} \\
& =& {\bf H_{k}} { \bf  \tilde x_{k}}^{-}+{\bf v_{k}} .
\end{array}
\label{innovation_error}
\end{eqnarray}
 The expectation of innovation error \eqref{innovation_error} is zero and its covariance is

  \begin{eqnarray*}
  \begin{array}{lcl}
E [{\bf e_{k}}{\bf e_{k}^{T}}]& =&  E [ ({\bf H_{k}} { \bf  \tilde x_{k}}^{-}+{\bf v_{k}})  ({\bf H_{k}} { \bf  \tilde x_{k}}^{-}+{\bf v_{k}})^{\top}]\\
& =&E[{\bf H_{k}} { \bf  \tilde x_{k}}^{-}{ \bf  \tilde x_{k}}^{-\top}{\bf H_{k}}^{\top}+ {\bf H_{k}} { \bf  \tilde x_{k}}^{-}{\bf v_{k}}^{\top}+\\&&{\bf v_{k}}{ \bf  \tilde x_{k}}^{-\top} {\bf H_{k}}^{\top}+{\bf v_{k}} {\bf v_{k}}^{\top}]\\
& = &{\bf H_{k}} E[{ \bf  \tilde x_{k}}^{-}{ \bf  \tilde x_{k}}^{-\top}] {\bf H_{k}}^{\top}+ E[{\bf v_{k}} {\bf v_{k}}^{\top}]\\
& = &{\bf H_{k}} {\bf \hat P_{k}}^{-} {\bf H_{k}}^{\top} + {\bf R_{k}}\\
&=&\sum\limits_{i=1}^{2L} {({\bf Z_{k-1}})_i} {({\bf Z_{k-1}})_i^\top}-{\bf \hat z_{k-1}}^{-} {\bf \hat z_{k-1}}^{{-}^\top} + {\bf R_{k}}\\
&=&{\bf P_{zz,k}}.
\end{array}
\end{eqnarray*}
 The predicted error is
    \begin{eqnarray}
    \begin{array}{lcl}
{ \bf  \tilde x_{k}}^{-}&= & {\bf {  x_{k}}}-{ \bf \hat x_{k}^{-}}  \\
& =& f({ \bf  x_{k-1}})+ {\bf w_{k-1}}-f({ \bf \hat x_{k}^{-}}) \\
& = &{\bf F_{k}} ({ \bf x_{k-1}}-{ \bf \hat x_{k-1}}^{-})+{\bf w_{k-1}}\\
& = & {\bf F_{k}} ({\bf \tilde x_{k-1}}^{-})+{\bf w_{k-1}}.
\end{array}
\label{approx}
\end{eqnarray}
Due to the classical approximation of equation \eqref{approx},  the noise sequence ${\bf w_{k-1}}$ can be changed to exact equality  as
 \begin{equation}
  { \bf  \tilde x_{k}}^{-}   \cong {\beta_{k-1}} {\bf F_{k}}{\bf \tilde x_{k-1}}^{-}.
  \label{beta}
 \end{equation}
Now, we take residual account at each iteration, in order to obtain an exact equality, we introduced unknown time-varying diagonal matrix ${\bf \beta_{k}}$ as
 \begin{eqnarray}
 { \bf  \tilde x_{k}}^{-} &\cong& {\bf \zeta_{k-1}} {\bf F_{k}}{ \bf  \tilde x_{k-1}}^{-}
\label{digonal_matrices}
  \end{eqnarray}

Next,  from (24), we have ${\bf \hat {x}_{k}}-{\bf {x}_{k}}=  {\bf \hat {x}_{k}}^{-}+{\bf K_{k}}{\bf e_{k}}-{\bf {x}_{k}}$ and further

\begin{align}
{ \bf  \tilde x_{k}} & = ( { \bf  \tilde x_{k}}^{-} -( \frac{1}{\bf \alpha_{k}}{\bf \hat P_{k}}^{-} {\bf H^\top} [\frac{1}{\bf \alpha_{k}}{\bf H} {\bf \hat P_{k}}^{-}{ \bf H^\top}+{\bf  {\hat R}_{k}}]^{-1}){\bf e
_{k}} ).
\end{align}

Now, we can use the auto-covariance of state and measurements are 
 \begin{equation}
\frac{1}{\bf \alpha_{k}}{\bf \hat P_{k}}{ \bf H^\top}{\bf \hat R_{k}^{-1}}= \frac{1}{\bf \alpha_{k}}{\bf P_{xz,k}}[\frac{1}{\bf \alpha_{k}}{\bf P_{zz,k}}+{\bf \hat R_{k}}]^{-1}
\label{egn:30}
 \end{equation}

  and equation (60)  becomes
       \begin{eqnarray}
{ \bf  \tilde x_{k}} & =  { \bf  \tilde x_{k}}^{-} - \frac{1}{\bf \alpha_{k}}{\bf \hat P_{k}} {\bf H^\top}{\bf \hat R_{k}^{-1}}{\bf e_{k}}
\label{mod_error}
\end{eqnarray}
while
   \begin{equation}
{\bf  \hat{P}_{k}^{-1}}= {\bf \alpha_{k}}({\bf \hat P_{k}}^{-})^{-1}+{\bf H^\top}{\bf \hat R_{k}^{-1}} {\bf H}.
\label{Pk}
 \end{equation}
 Substituting \eqref{mod_error} into \eqref{Lyap}, it follows
  \begin{align}
  \begin{split}
 { \bf V_{k} } = &({ \bf  \tilde x_{k}}^{-}-{\frac{1}{\bf \alpha_{k}}}{\bf P_{xz,k}}{\bf \hat R_{k}^{-1}}{\bf e_{k}})^{\top}  {\bf \hat P_{k}}^{-1}\\
  &({ \bf  \tilde x_{k}}^{-}-{ \frac{1}{\bf \alpha_{k}}}{\bf P_{xz,k-1}}  {\bf \hat R_{k}^{-1}} {\bf e_{k}})\\
= &{\bf \tilde x_{k}^{-\top}}{\bf \hat{P}_{k}^{-1}}{\bf \tilde x_{k}} +{\frac{1}{\bf \alpha_{k}^{2}}}{\bf e_{k}}^{\top} {\bf \hat R_{k}^{-1}}{\bf P_{zz,k}} {\bf \hat R_{k}^{-1}}{\bf e_{k}} \\
& -\frac{2}{\alpha_k} {\bf \tilde x_{k}^{-\top}}{ \bf H^\top}{\bf \hat R_{k}^{-1}}{\bf e_{k}}\\
 =&{\bf \alpha_{k}}{\bf \tilde x_{k}^{-\top}}({\bf \hat P_{k}}^{-})^{-1}{\bf \tilde x_{k}^{-}}+ {\bf \tilde x_{k}^{-\top}}{\bf H^\top}{\bf \hat R_{k}}^{-1} {\bf H}{\bf \tilde x_{k}^{-}} \\
  &+ {\frac{1}{\bf \alpha_{k}^{2}}}{\bf e_{k}}^{\top} {\bf \hat  R_{k}^{-1}}{\bf P_{zz,k}} {\bf \hat R_{k}^{-1}}{\bf e_{k}} \\&-\frac{2}{\alpha_k} {\bf \tilde x_{k}^{-\top}}{ \bf H^\top}{\bf  \hat R_{k}^{-1}}{\bf e_{k}}\\
   =&{\bf \alpha_{k}} {\bf V_{k}^{-}}+ {\bf \tilde x_{k}^{-\top}}{\bf H^\top}{\bf \hat R_{k}}^{-1} {\bf H}{\bf \tilde x_{k}^{-}} \\
  &+ {\frac{1}{\bf \alpha_{k}^{2}}}{\bf e_{k}}^{\top} {\bf \hat R_{k}^{-1}}{ \bf H} {\bf P_{xz,k-1}} {\bf \hat R_{k}^{-1}}{\bf e_{k}} \\&-\frac{2}{\alpha_k} {\bf \tilde x_{k}^{-\top}}{ \bf H^\top}{\bf  \hat R_{k}^{-1}}{\bf e_{k}}
\end{split}
\label{lyap_deriv}
\end{align}
where $ {\bf V_{k}^{-}}={\bf \tilde x_{k}^{-\top}}({\bf \hat P_{k}}^{-})^{-1}{\bf \tilde x_{k}^{-}}$. Using \eqref{beta}  and the identity
\begin{align*}
 {\bf \hat P_{k}}^{-}=\frac{1}{2L}\sum\limits_{i=1}^{2L}{({\bf X_{k}})_i} {({\bf X_{k}})_i^\top} -{\bf \hat x_{k}}{\bf \hat x_{k}}^{{-}^\top}+{\bf Q_{k}}
\end{align*}
we have
\begin{align}
\begin{split}
{\bf V_{k}^{-}}=&{ \bf \alpha_{k}} {\bf \tilde{x}_{k-1}^\top}({\bf A^{\top}}  {\bf \zeta_{k-1}} (\frac{1}{2L}\sum\limits_{i=1}^{2L}{({\bf X_{k}})_i} {({\bf X_{k}})_i^\top} \\&-{\bf \hat x_{k}}{\bf \hat x_{k}}^{{-}^\top}+{\bf Q_{k}})^{-1} \times  {\bf \zeta_{k-1}}  {\bf A}) {\bf \tilde{x}_{k-1}}
\end{split}
\end{align}
and \eqref{lyap_deriv} becomes
\begin{align}
\begin{split}
 { \bf V_{k} }= &  {\bf V_{k}^{-}}+{\bf e_{k}}^{\top}(   \frac{1}{\bf  \alpha_{k}^{2}}{\bf \hat  R_{k}^{-1}}[\frac{1}{2L}\sum\limits_{i=1}^{2L} {({\bf Z_{k-1}})_i} {({\bf Z_{k-1}})_i^\top}\\&-{\bf \hat z_{k-1}}^{-} {\bf \hat z_{k-1}}^{{-}^\top}]{\bf \hat R_{k}^{-1}}
 +(1 -\frac{2}{\bf \alpha_k}) {\bf \hat R_{k}^{-1}}){\bf e_{k}}
 \end{split}
 \end{align}
 A decreasing sequence  $ \{{\bf V_{k}\}_{k=1}^{\infty}}$ implies
   \begin{equation}
 {\bf V_{k}}-{ \bf V_{k-1} }={ \bf V_{k} }-{ \bf V_{k}^{-} }+{ \bf V_{k}^{-} }-{ \bf V_{k-1} } \le 0.
\label{V_cond}
  \end{equation}
Thus
\begin{align}\label{eqn:34}
 \begin{split}
 { \bf V_{k} }-{ \bf V_{k-1} }= &{\bf e_{k}}^{\top} [(1 -\frac{2}{\bf \alpha_k}) {\bf \hat R_{k}^{-1}}\\& + {\frac{1}{\bf \alpha_{k}^{2}}} {\bf \hat {R}_{k}^{-1}}[{ \bf H} {\bf \hat P_{k}} { \bf H^\top}]{\bf \hat {R}_{k}^{-1}}] {\bf e_{k}} 
 \\& + {\bf\tilde{x}_{k-1}}^{-\top}( {\bf \alpha_{k}}{\bf F_{k}^{\top}} {\bf \zeta_{k}}(\frac{1}{2L}\sum\limits_{i=1}^{2L}{({\bf X_{k}})_i} {({\bf X_{k}})_i^\top} \\&-{\bf \hat x_{k}}{\bf \hat x_{k}}^{{-}^\top}+{\bf Q_{k}})^{-1} {\bf \zeta_{k}} {\bf F_{k}}-{\bf \hat P_{k-1}^{-1}} ) {\bf \tilde x_{k-1}}  \le 0.
\end{split}
\end{align}
in view of the inequality  \eqref{condition}.
\end{proof}
%


\bibliographystyle{ieeetr}
\bibliography{Reference}

\end{document}